\documentclass[twoside,slac_one]{revtex4}
\usepackage{graphicx}
\usepackage{fancyhdr}
\usepackage{amsmath} 
\usepackage{bm}
\usepackage{amsxtra}
\usepackage{amssymb}
\usepackage{amsthm}
\usepackage{latexsym}
\usepackage{lscape}

\begin{document}

\title{After LUX: The LZ Program}

%

\author{D.C. Malling} 
 \email{David_Malling@brown.edu}
 \affiliation{Brown University, Dept. of Physics, 182 Hope St., Providence RI 02912, USA}

\author{D.~S. Akerib} 
 \affiliation{Case Western Reserve University, Dept. of Physics, 10900 Euclid Ave, Cleveland OH 44106, USA}
 
\author{H.M.~Ara\'{u}jo} 
 \affiliation{High Energy Physics group, Blackett Laboratory, Imperial College London, UK}

\author{X. Bai} 
 \affiliation{South Dakota School of Mines and technology, 501 East St Joseph St., Rapid City SD 57701, USA}

\author{S. Bedikian} 
 \affiliation{Yale University, Dept. of Physics, 217 Prospect St., New Haven CT 06511, USA}

\author{E. Bernard} 
 \affiliation{Yale University, Dept. of Physics, 217 Prospect St., New Haven CT 06511, USA}

\author{A. Bernstein} 
 \affiliation{Lawrence Livermore National Laboratory, 7000 East Ave., Livermore CA 94551, USA}

\author{A. Bradley} 
 \affiliation{Case Western Reserve University, Dept. of Physics, 10900 Euclid Ave, Cleveland OH 44106, USA}

\author{S.B. Cahn} 
 \affiliation{Yale University, Dept. of Physics, 217 Prospect St., New Haven CT 06511, USA}

\author{M.C. Carmona-Benitez} 
 \affiliation{Case Western Reserve University, Dept. of Physics, 10900 Euclid Ave, Cleveland OH 44106, USA}

\author{D. Carr} 
 \affiliation{Lawrence Livermore National Laboratory, 7000 East Ave., Livermore CA 94551, USA}
 
\author{J.J. Chapman} 
 \affiliation{Brown University, Dept. of Physics, 182 Hope St., Providence RI 02912, USA}

\author{K. Clark} 
 \affiliation{Case Western Reserve University, Dept. of Physics, 10900 Euclid Ave, Cleveland OH 44106, USA}

\author{T. Classen} 
 \affiliation{University of California Davis, Dept. of Physics, One Shields Ave., Davis CA 95616, USA}

\author{T. Coffey} 
\affiliation{Case Western Reserve University, Dept. of Physics, 10900 Euclid Ave, Cleveland OH 44106, USA}

\author{A.~Currie} 
 \affiliation{High Energy Physics group, Blackett Laboratory, Imperial College London, UK}

\author{S. Dazeley} 
 \affiliation{Lawrence Livermore National Laboratory, 7000 East Ave., Livermore CA 94551, USA}

\author{L. de~Viveiros} 
 \affiliation{LIP-Coimbra, Department of Physics, University of Coimbra, Rua Larga, 3004-516 Coimbra, Portugal}

\author{M. Dragowsky} 
 \affiliation{Case Western Reserve University, Dept. of Physics, 10900 Euclid Ave, Cleveland OH 44106, USA}

\author{E. Druszkiewicz} 
 \affiliation{University of Rochester, Dept. of Physics and Astronomy, Rochester NY 14627, USA}

\author{C.H. Faham} 
 \affiliation{Brown University, Dept. of Physics, 182 Hope St., Providence RI 02912, USA}

\author{S. Fiorucci} 
 \affiliation{Brown University, Dept. of Physics, 182 Hope St., Providence RI 02912, USA}

\author{R.J. Gaitskell} 
 \affiliation{Brown University, Dept. of Physics, 182 Hope St., Providence RI 02912, USA}

\author{K.R. Gibson} 
 \affiliation{Case Western Reserve University, Dept. of Physics, 10900 Euclid Ave, Cleveland OH 44106, USA}

\author{C. Hall} 
 \affiliation{University of Maryland, Dept. of Physics, College Park MD 20742, USA}

\author{M. Hanhardt} 
 \affiliation{South Dakota School of Mines and technology, 501 East St Joseph St., Rapid City SD 57701, USA}

\author{B. Holbrook} 
 \affiliation{University of California Davis, Dept. of Physics, One Shields Ave., Davis CA 95616, USA}

\author{M. Ihm} 
 \affiliation{University of California Berkeley, Department of Physics, Berkeley, CA 94720-7300, USA}

\author{R.G. Jacobsen} 
 \affiliation{University of California Berkeley, Department of Physics, Berkeley, CA 94720-7300, USA}

\author{L. Kastens} 
 \affiliation{Yale University, Dept. of Physics, 217 Prospect St., New Haven CT 06511, USA}

\author{K. Kazkaz} 
 \affiliation{Lawrence Livermore National Laboratory, 7000 East Ave., Livermore CA 94551, USA}

\author{R. Lander} 
 \affiliation{University of California Davis, Dept. of Physics, One Shields Ave., Davis CA 95616, USA}

\author{N. Larsen} 
 \affiliation{Yale University, Dept. of Physics, 217 Prospect St., New Haven CT 06511, USA}

\author{C. Lee} 
 \affiliation{Case Western Reserve University, Dept. of Physics, 10900 Euclid Ave, Cleveland OH 44106, USA}

\author{D. Leonard} 
 \affiliation{University of Maryland, Dept. of Physics, College Park MD 20742, USA}

\author{K. Lesko} 
 \affiliation{Lawrence Berkeley National Laboratory, 1 Cyclotron Rd., Berkeley CA 94720, USA}
 
\author{A. Lindote}
 \affiliation{LIP-Coimbra, Department of Physics, University of Coimbra, Rua Larga, 3004-516 Coimbra, Portugal}

\author{M.\,I. Lopes}
 \affiliation{LIP-Coimbra, Department of Physics, University of Coimbra, Rua Larga, 3004-516 Coimbra, Portugal}

\author{A. Lyashenko} 
 \affiliation{Yale University, Dept. of Physics, 217 Prospect St., New Haven CT 06511, USA}

\author{P.~Majewski} 
 \affiliation{Particle Physics Department, STFC Rutherford Appleton Laboratory, Chilton, UK}

\author{R. Mannino} 
 \affiliation{Texas A \& M University, Dept. of Physics, College Station TX 77843, USA}

\author{D.N. McKinsey} 
 \affiliation{Yale University, Dept. of Physics, 217 Prospect St., New Haven CT 06511, USA}

\author{D.-M Mei} 
 \affiliation{University of South Dakota, Dept. of Physics, 414E Clark St., Vermillion SD 57069, USA}

\author{J. Mock} 
 \affiliation{University of California Davis, Dept. of Physics, One Shields Ave., Davis CA 95616, USA}

\author{M. Morii} 
 \affiliation{Harvard University, Dept. of Physics, 17 Oxford St., Cambridge MA 02138, USA}

\author{A.St\,J.~Murphy} 
 \affiliation{School of Physics \& Astronomy, University of Edinburgh, UK}

\author{H. Nelson} 
 \affiliation{University of California Santa Barbara, Dept. of Physics, Santa Barbara, CA, USA}

\author{F. Neves}
 \affiliation{LIP-Coimbra, Department of Physics, University of Coimbra, Rua Larga, 3004-516 Coimbra, Portugal}

\author{J.A. Nikkel} 
 \affiliation{Yale University, Dept. of Physics, 217 Prospect St., New Haven CT 06511, USA}

\author{M. Pangilinan} 
 \affiliation{Brown University, Dept. of Physics, 182 Hope St., Providence RI 02912, USA}

\author{P. Phelps} 
 \affiliation{Case Western Reserve University, Dept. of Physics, 10900 Euclid Ave, Cleveland OH 44106, USA}

\author{L.~Reichhart} 
 \affiliation{School of Physics \& Astronomy, University of Edinburgh, UK}

\author{T. Shutt} 
 \affiliation{Case Western Reserve University, Dept. of Physics, 10900 Euclid Ave, Cleveland OH 44106, USA}

\author{C. Silva}
 \affiliation{LIP-Coimbra, Department of Physics, University of Coimbra, Rua Larga, 3004-516 Coimbra, Portugal}

\author{W. Skulski} 
 \affiliation{University of Rochester, Dept. of Physics and Astronomy, Rochester NY 14627, USA}

\author{V. Solovov}
 \affiliation{LIP-Coimbra, Department of Physics, University of Coimbra, Rua Larga, 3004-516 Coimbra, Portugal}

\author{P. Sorensen} 
 \affiliation{Lawrence Livermore National Laboratory, 7000 East Ave., Livermore CA 94551, USA}

\author{J. Spaans} 
 \affiliation{University of South Dakota, Dept. of Physics, 414E Clark St., Vermillion SD 57069, USA}

\author{T. Stiegler} 
 \affiliation{Texas A \& M University, Dept. of Physics, College Station TX 77843, USA}

\author{T.J.~Sumner} 
 \affiliation{High Energy Physics group, Blackett Laboratory, Imperial College London, UK}

\author{R. Svoboda} 
 \affiliation{University of California Davis, Dept. of Physics, One Shields Ave., Davis CA 95616, USA}

\author{M. Sweany} 
 \affiliation{University of California Davis, Dept. of Physics, One Shields Ave., Davis CA 95616, USA}

\author{M. Szydagis} 
 \affiliation{University of California Davis, Dept. of Physics, One Shields Ave., Davis CA 95616, USA}

\author{J. Thomson} 
 \affiliation{University of California Davis, Dept. of Physics, One Shields Ave., Davis CA 95616, USA}

\author{M. Tripathi} 
 \affiliation{University of California Davis, Dept. of Physics, One Shields Ave., Davis CA 95616, USA}

\author{J.R. Verbus} 
 \affiliation{Brown University, Dept. of Physics, 182 Hope St., Providence RI 02912, USA}

\author{N. Walsh} 
 \affiliation{University of California Davis, Dept. of Physics, One Shields Ave., Davis CA 95616, USA}

\author{R. Webb} 
 \affiliation{Texas A \& M University, Dept. of Physics, College Station TX 77843, USA}

\author{J.T. White} 
 \affiliation{Texas A \& M University, Dept. of Physics, College Station TX 77843, USA}

\author{M. Wlasenko} 
 \affiliation{Harvard University, Dept. of Physics, 17 Oxford St., Cambridge MA 02138, USA}

\author{F.L.H. Wolfs} 
 \affiliation{University of Rochester, Dept. of Physics and Astronomy, Rochester NY 14627, USA}

\author{M. Woods} 
 \affiliation{University of California Davis, Dept. of Physics, One Shields Ave., Davis CA 95616, USA}

\author{C. Zhang} 
 \affiliation{University of South Dakota, Dept. of Physics, 414E Clark St., Vermillion SD 57069, USA}

\begin{abstract}
The LZ program consists of two stages of direct dark matter searches using liquid Xe detectors. The first stage will be a 1.5-3~tonne detector, while the last stage will be a 20~tonne detector. Both devices will benefit tremendously from research and development performed for the LUX experiment, a 350~kg liquid Xe dark matter detector currently operating at the Sanford Underground Laboratory. In particular, the technology used for cryogenics and electrical feedthroughs, circulation and purification, low-background materials and shielding techniques, electronics, calibrations, and automated control and recovery systems are all directly scalable from LUX to the LZ detectors. Extensive searches for potential background sources have been performed, with an emphasis on previously undiscovered background sources that may have a significant impact on tonne-scale detectors. The LZ detectors will probe spin-independent interaction cross sections as low as $5\times10^{-49}$~cm$^2$ for 100~GeV WIMPs, which represents the ultimate limit for dark matter detection with liquid xenon technology.
\end{abstract}

\maketitle

\thispagestyle{fancy}




\section{Introduction}
Observational evidence from the past 80 years strongly supports the theory of a non-luminous matter component which comprises 80\% of all matter in the universe. The primary candidate for this dark matter particle is the weakly interacting massive particle (WIMP). Direct dark matter searches seek to detect the presence of dark matter through detection of weak interactions between WIMPs and atomic nuclei.

Liquid xenon time projection chambers (TPCs) are a proven type of direct dark matter detector design. These detectors use photosensors (typically photomultiplier tubes, or PMTs) at the top and bottom of the detector volume to read out scintillation signals resulting from Xe nuclear or electronic recoils. An electric field drifts the resulting ionization from the recoil out of the liquid Xe volume and into a gaseous region above, where the electrons produce an electroluminescence signal. By measurement of both signals, these detectors make a measurement of the event recoil energy. The relative sizes of the scintillation and ionization yields are used to discriminate between nuclear recoils (NR) and electron recoils (ER) (the primary background for these detectors). Detector volume fiducialization, an important background rejection technique, is accomplished by analysis of the ionization signal hit pattern concentration within the photosensors for XY positioning, as well as the timing between scintillation and ionization signals for Z positioning.

\begin{figure}[ht]
\centering
\includegraphics[width=135mm]{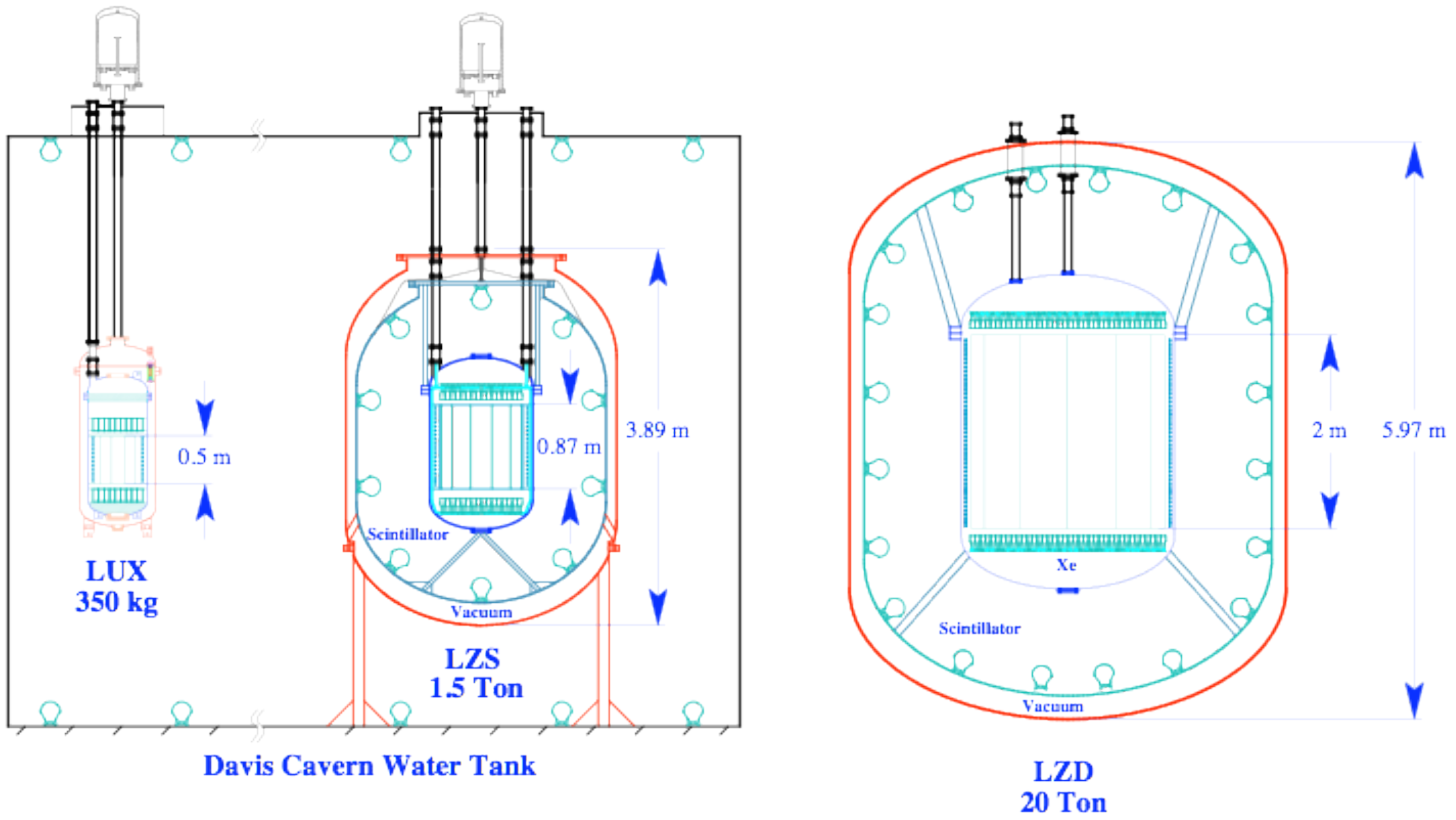}
\caption{Size comparison of the LUX, LZS, and LZD detectors, including LZ scintillator volumes. LZS and LZD represent mass scale-ups beyond LUX by factors of $\times$5-10 and $\times$60, respectively. The modest increase in linear dimension for the LZS active region allows the detector to make use of the existing LUX water tank. LZD will require a 12~m $\times$ 12~m water shield in a separate lab space.} \label{fig:lux-lz-comparison}
\end{figure}

The LZ program (LUX-ZEPLIN) focuses on the construction of two liquid Xe TPC detectors at the Sanford Underground Research Facility, or SURF. The first detector, named LZS \cite{LZSProposal2009}, will use 1.5 to 3~tonnes of liquid Xe as its target mass. After successfully running this detector, a 20~tonne detector, named LZD \cite{LZDProposal2009}, will be constructed. As will be discussed, this is the largest liquid Xe TPC dark matter detector that can be built before irreducible backgrounds are encountered. The LZ detector cross sections are drawn to scale in comparison to LUX in Fig. \ref{fig:lux-lz-comparison}.

The LZ detectors will greatly benefit from research and development performed for the LUX experiment \cite{LUXProposal2007,LUXwebsite}. The LZS detector has been designed such that it will be able to be deployed in the existing LUX laboratory space at the Homestake 4850~ft level, and will even be able to effectively use the existing LUX water shield for suppression of external backgrounds. Many LZ technologies benefit from research and development on LUX systems. A review of LZ technologies is described in Sec. \ref{sec:LZ-Technologies}.

Liquid Xe TPCs face a background composed primarily of gamma rays generated from natural radioactive decay, as well as neutrons from both radioactive decay and muon interactions in cavern rock and detector materials. As with LUX, the LZ detectors will reduce backgrounds from external sources (primarily cavern rock) through the use of a water shield. Internal backgrounds will be moderated through a rigorous material selection and screening program. Consideration has also been given to examination of backgrounds which had not previously been relevant for much smaller liquid Xe TPC experiments. Background models and expectations are discussed in Sec. \ref{sec:Backgrounds}.

\section{LZ Technologies}
\label{sec:LZ-Technologies}

\subsection{PMTs}
\label{sub:PMTs}

\begin{figure}[ht]
\centering
\includegraphics[width=80mm]{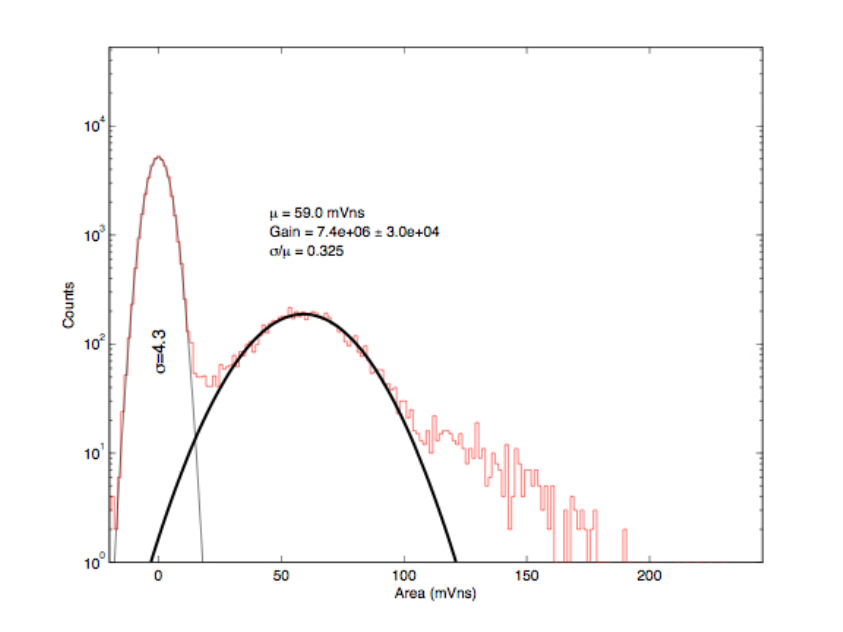}
\caption{Single photoelectron and gain measurements with a 7.6~cm Hamamatsu R11065 PMT. The PMT was tested in liquid xenon (-100$^\circ$~C) at 1500~V. These tubes are shown to perform comparably in liquid xenon as compared to the 5.7~cm R8778 PMTs used in LUX.} \label{fig:R11065-testing}
\end{figure}

LZ will use $\sim$200 and $\sim$1000 7.6~cm PMTs for instrumentation of the LZS and LZD active regions, respectively. These PMTs are very similar in construction to the 5.7~cm diameter Hamamatsu R8778 PMTs used and extensively tested by LUX, but with a photocathode area a factor of $\times$2 larger. Prototype Hamamatsu 7.6~cm diameter R11065 PMTs have been tested in liquid Xe, and have been found to perform comparably to the LUX R8778 PMTs (Fig. \ref{fig:R11065-testing}).

PMTs generally represent a major radioactive background source for liquid Xe TPCs. This is due to both the complexity of their construction and their proximity to the active region. A screening program to identify ultra low-background PMTs has been undertaken for LZ construction, with a goal towards bringing the PMT background contributions subdominant to other background sources (see Sec. \ref{sec:Backgrounds}). The leading candidate for use in LZ is the Hamamatsu R11410~MOD, identical to the R11065 but utilizing extremely low-radioactivity components. This 7.6~cm PMT yields 90\% counting limits of $<$0.4~$^{238}$U / $<$0.3~$^{232}$Th mBq/PMT, representing a combined factor of $\times$1/15 in gamma emission and $\times$1/18 in neutron emission as compared to the R8778 PMTs used in LUX.

\subsection{Cryogenics, Circulation, and Purification}

\begin{figure}[ht]
\centering
\includegraphics[width=80mm]{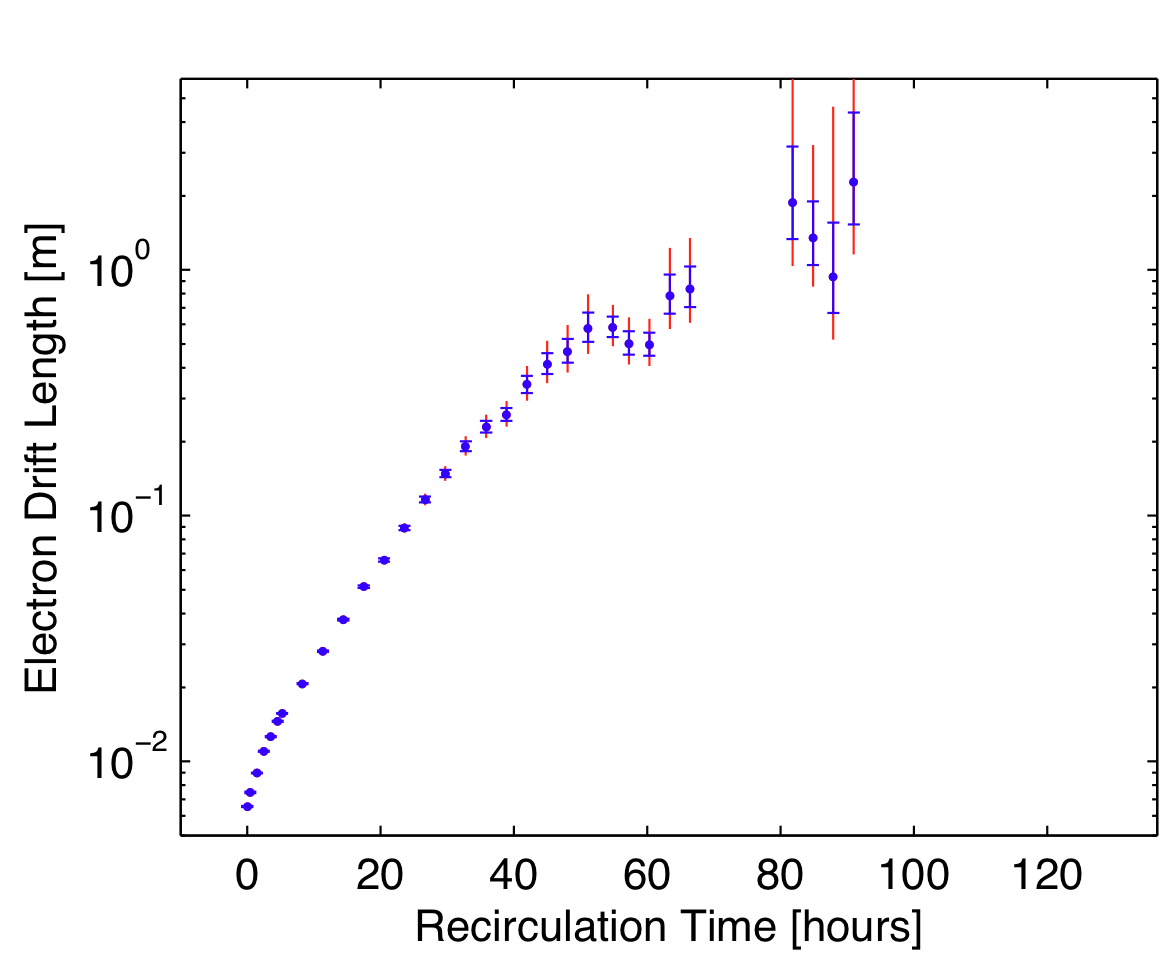}
\caption{Electron drift length in liquid Xe over time in the LUX~0.1 experiment. 1$\sigma$ and 2$\sigma$ error bars are shown for each measurement. 55~kg Xe were circulated and purified while testing final designs for the LUX heat exchanger and weir. The initial purification time constant was measured at 9$\pm$1~hrs, unprecedented for a detector with a thermal mass comparable to LUX~0.1. The system will be tested with LUX during initial running, and is directly scalable for the LZ detectors.} \label{fig:lux01-purification}
\end{figure}

The cooling and Xe circulation systems used in LUX were extensively tested with the LUX~0.1 prototype. This LUX prototype consisted of 55~kg of Xe and 270~kg of Al displacer, as well as a fully-functional central TPC with 0.5~kg of liquid Xe and four R8778 PMTs, all housed within a stainless steel double-walled cryostat built to contain the full LUX experiment. Using a novel heat exchanger design which has since been implemented in LUX, the cooling power required for the detector was reduced from 500~W to 20~W. This heat exchanger, coupled with a circulation system that uses a weir to set the liquid height and a commercial gas-phase getter for purification, produced a $>$1~m electron drift length (the greatest length measurable with the 5~cm high TPC) on a time scale of $\sim$4 days, as shown in Fig. \ref{fig:lux01-purification}.

The LUX and LZ detectors will be cooled by thermosyphons, comprised of closed loops of N$_2$ which efficiently cycle heat from the detector internals to a liquid N$_2$ reservoir. Each main LUX thermosyphon is capable of delivering $>$1~kW of cooling power, while the projected power required for cooling the detector itself is of order 10~W. This means that the LUX thermosyphon system can easily be scaled and used for LZ, with no significant improvements in technology required.

LUX will test a charcoal-based system for the removal of Kr traces from Xe. Of particular concern is $^{85}$Kr, a beta emitter with an endpoint at 687~keV, which can contribute significantly to internal backgrounds (Sec. \ref{sub:Intrinsic-Backgrounds}). The LUX system will be able to purify Xe at levels $<$5~ppt, sufficient for subdominance of the $^{85}$Kr signature for WIMP searches. This technology is directly scalable for the LZ detectors, with orders of magnitude improvement possible.

\subsection{Internals and Material Selection}

Through the use of a large water shield instrumented with an active veto, external backgrounds have been rendered subdominant to background goals for the LUX experiment. From background modeling work for LZ, internal radioactivity is expected to be the most significant source of backgrounds for these detectors as well. In order to maintain low backgrounds from internal components, major internal components are created using materials known to have high radiopurity, e.g. OFHC Cu, HDPE, and PTFE. A radiopurity screening program, similar to that used for LUX, will be implemented for selection of materials for all major internal items, ensuring that internal backgrounds are limited below goals.

The LUX collaboration identified several highly radiopure samples of Ti ($<$0.4~mBq/kg $^{238}$U, $<$0.8~mBq/kg $^{232}$Th, $<$1.6~mBq/kg $^{40}$K), enabling the use of that material for construction of the inner and outer cryostat vessels. Ti is a very attractive option for cryostat construction due to its strength, cleanliness, and relative transparency to gamma rays, in comparison with steel and Cu which are typically used for construction of cryogenic vessels. Gamma transparency is attractive when using an external scintillator veto, which can catch escaping gammas to provide rejection of some intrinsic backgrounds (Sec. \ref{sub:Scintillator}).

Under investigation is the use of ``active internals,'' which would be separately-instrumented scintillator plastics used in place of Cu or HDPE in the construction of several of the major internal components. Though OFHC Cu is extremely radiopure, it can also shield gammas from scattering in the external scintillator veto. The projected background reduction factor from using scintillating internals is currently under investigation.

An addition of particular interest is the use of a fiber optic camera in the interior of the detector. This will enable direct monitoring of liquid surface conditions, sparking events, and grid condition. A prototype of the LZ camera system is being prepared for use in a future data run in LUX.

\subsection{Electronics}

LUX currently uses a VME-based acquisition system with 128 separate channels, with a custom-built 16-channel FPGA-based DDC-8 trigger system. The trigger system, extremely flexible in terms of implementation of pulse analysis algorithms, and allowing advanced data cuts to take place before pulse acquisition, can easily be expanded to hundreds of channels. For the LZ detectors, the DDC-8 system will be expanded to provide separate channels for all individual PMTs. The same digitizers will also provide data acquisition functionality, replacing the VME-based LUX DAQ system.

\subsection{Calibrations}

The self-shielding properties of liquid Xe are extremely advantageous in the rejection of gamma backgrounds. Unfortunately, the self-shielding makes it extremely difficult to calibrate the detector with external gamma sources. For liquid Xe TPCs the size of LUX or greater, external calibration sources are grossly inefficient at producing low-energy ER events within the fiducial volume. For this reason, internal calibration sources are being developed and deployed for testing in LUX, with an eye toward adaptation for the LZ detectors. These sources will maximize calibration efficiency at the center of the detector for both fixed- and variable-energy calibrations, which will set the light yield scale at the WIMP search region and determine ER rejection capability, respectively.

For fixed-energy calibrations, $^{83}$Kr$^{\text{m}}$ is a favored candidate \cite{Kastens2009,Kastens2010}. $^{83}$Kr$^{\text{m}}$ generates two conversion electrons at 32.1 and 9.4~keV, separated by a characteristic time of 154~ns. The characteristic time separation is long enough to be easily distinguished during analysis. The 1.8~hr half-life of $^{83}$Kr$^{\text{m}}$ quickly reduces the source activity below appreciable levels on the time scale of days, and the signature of the isotope allows any residual contamination to be easily rejected during dark matter searches.

Variable-energy calibrations can be accomplished with a beta emitter. One particular candidate under investigation is tritium, with a Q-value of 18.6~keV. Tritium would be introduced in the form of tritiated methane. Removal of tritiated methane at ppm levels from gaseous Xe has been achieved with efficiencies exceeding 99.98\% \cite{Dobi2010}. Significant advances in purification must be attained before use in liquid xenon, as background subdominance of the tritiated methane beta signature requires it to be present at levels $<$10$^{-25}$~g/g in Xe in LUX, and $<$10$^{-28}$~g/g in LZD, during WIMP searches.

\subsection{Scintillator}
\label{sub:Scintillator}

A major addition to the LZ detectors is the presence of an organic scintillator volume surrounding the active region. The scintillator will be instrumented with large-area PMTs, and will serve primarily to enhance rejection of internal backgrounds, as discussed in Sec. \ref{sub:Internal-Backgrounds}. The scintillator will likely be operated at liquid Xe temperatures to enhance light yields \cite{Britvich1999}, though development work remains to be completed to determine the impact of temperature on scintillator transparency, as well as explore an optimal cooling strategy and additional cryogenic safety concerns.

\subsection{Water Shield}

The LZ detectors will use a water shield to eliminate external backgrounds in a configuration similar to that of LUX. The LUX 8~m $\times$ 6~m water shield is equipped with 20 large-area PMTs which allow the water shield to provide Cherenkov veto functionality. The LZS detector will use the existing LUX water shield, as the LZS active region represents a modest scale-up in linear dimensions, and the layer of scintillator surrounding the active region will greatly supplement background reduction. LZD will use a 12~m $\times$ 12~m water shield with comparable or improved Cherenkov veto efficiency, which reduces external backgrounds far below significance (Sec. \ref{sub:External-Backgrounds}).

\section{Backgrounds}
\label{sec:Backgrounds}

Liquid Xe detectors greatly benefit from increases in mass, as the efficiency of their self-shielding capability dramatically increases as well. For a gamma or neutron to create a false WIMP signature, the particle must pass several cm into the detector, scatter through a small angle corresponding to a low energy deposition, and escape the detector without scattering again. The probability of escape drops exponentially with increases to linear detector dimension.

The increased size of the LZ detectors (50~cm $\times$ 50~cm for LUX; 85~cm $\times$ 85~cm or 1~m $\times$ 1~m for LZS 1.5 or 3~tonne respectively; 2~m $\times$ 2~m for LZD) will greatly aid rejection of backgrounds from both external and internal sources. However, background requirements for these detectors also become much more stringent as experiment running times increase. The increase in Xe mass and detector surface area also increases sensitivity to backgrounds which were not significant for smaller detectors. An extensive search has been performed to understand these background sources.

\subsection{External Backgrounds}
\label{sub:External-Backgrounds}

External backgrounds are generated primarily from muon spallation events in the rock, as well as muon capture in the detector water shield itself. Both event types produce a high-energy neutron flux which is able to penetrate to the fiducial volume of the detector with non-negligible efficiency. Extensive modeling work performed for the LUX experiment indicates that these background sources will remain subdominant to internal backgrounds. As LZS will use the existing LUX water shield and will offer enhanced self-shielding, these backgrounds are not considered to be significant for LZS either.

\begin{figure}[ht]
\centering
\includegraphics[width=80mm]{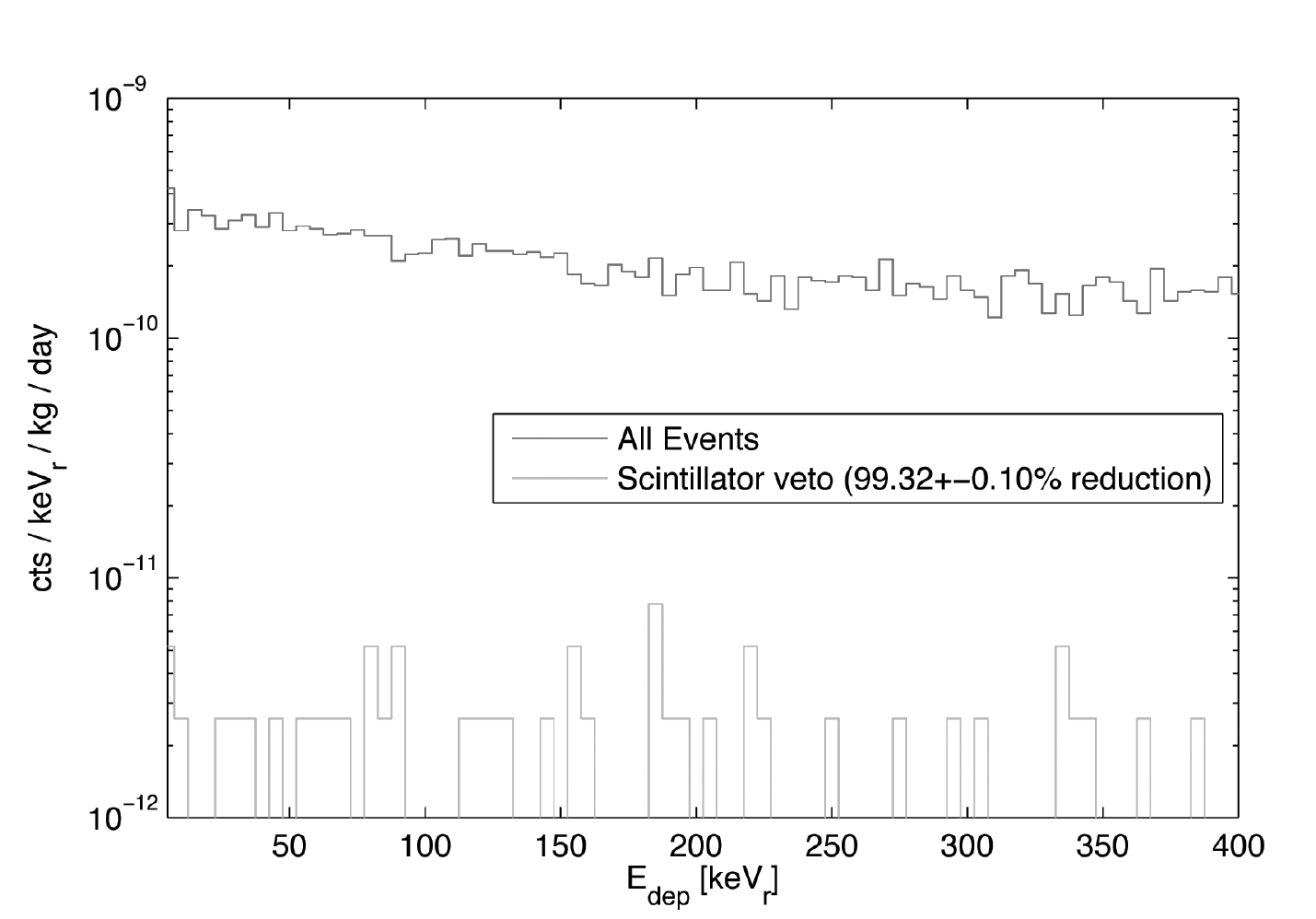}
\caption{Simulated nuclear recoil energy spectrum in LZD resulting from high-energy neutrons generated from muon spallation in cavern rock, before (dark) and after (light) application of a scintillator veto cut with a 100~keV threshold. The simulation assumed a 12~m $\times$ 12~m water tank with a 2~m $\times$ 2~m liquid xenon target at its center, surrounded by an additional 1~m thick liquid scintillator region (modeled as water). The initial neutron energy spectrum was obtained from \cite{Mei2005}. Within simulation statistics, standard fiducialization and single-scatter cuts eliminate all background events at $>$99.9\% efficiency. The simulation was performed using the LUXSim package.} \label{fig:lzd-rock-n-bg}
\end{figure}

Modeling of LZD (Fig. \ref{fig:lzd-rock-n-bg}), with a 12~m $\times$ 12~m water shield with 4~m radial water thickness and a 1~m thick scintillator shows that the neutron backgrounds generated from cavern rock are moderated to nearly an order of magnitude below PMT backgrounds, before application of analysis cuts. The simulation and subsequent analysis does not consider correlations with muon showers, which is expected to provide additional veto capability using muon Cherenkov light generated in the water shield. Early studies of neutrons generated within the water shield by muon capture show that they result in a rate of high-energy neutrons which is about half of the rate of neutrons from muon spallation in rock.

\subsection{Internal Backgrounds}
\label{sub:Internal-Backgrounds}

As with LUX, LZ backgrounds are expected to be limited by activity from internal components, in particular the PMTs. A rigorous material screening program will enforce this goal by combining detailed Monte Carlo modeling with counting results for construction materials used for all major internal components. The use of R11410~MOD PMTs (Sec. \ref{sub:PMTs}) significantly reduces the PMT contribution to the internal background. For LZD, the R11410~MOD PMTs will reduce PMT background contributions to a level an order of magnitude below the predicted background contribution from neutrino scattering (Sec. \ref{sub:Neutrino-Backgrounds}).

As the detector surface area increases, stringent precautions must be taken to limit exposure of detector internals to airborne Rn during construction. Rn progeny can plate out on detector surfaces and generate neutron backgrounds through ($\alpha$,n) interactions with the surrounding detector components, particularly fluorine-rich targets such as PTFE. As with LUX, care will be taken during the manufacturing process to keep all major internals under active N$_2$ purge in order to limit Rn exposure.

\subsection{Intrinsic Backgrounds}
\label{sub:Intrinsic-Backgrounds}

The increase in xenon mass for the LZ detectors brings a corresponding increase in sensitivity to intrinsic xenon backgrounds which were not considered for previous detectors. Of particular concern are a class of isotopes which undergo ``naked'' or ``semi-naked'' beta decay, meaning that the decay has a significant probability of emitting a beta without an accompanying gamma which would enhance veto capabilities. Naked beta decays produce no accompanying gamma, while semi-naked beta decays produce an accompanying gamma with enough energy to potentially escape the active region. One notable example of a naked beta emitter is $^{214}$Bi, a member of the $^{238}$U decay chain, which can be introduced into the xenon through Rn emanation from detector internals. Activation of Xe through muon capture, neutron capture, or fast neutron activation is also of concern.

The dominant intrinsic background has been found to be xenon isotopes generated by fast neutron activation and thermal neutron capture. The most important isotope is $^{137}$Xe, which has a 2/3 probability of emitting a naked beta with a 4.1~MeV endpoint, and a 1/3 chance of emitting a semi-naked 3.7~MeV beta in coincidence with a 450~keV gamma. This background source yields event rates approximately two orders of magnitude below those expected from neutrinos, as discussed in Sec. \ref{sub:Neutrino-Backgrounds}, and is thus not a concern for LZ.

Kr is present in commercial Xe in $\sim$ppb concentrations. $^{85}$Kr, a naked beta emitter, is present within natural Kr at concentrations of $\sim$10~ppt. As Kr is not removed by the getters planned for use in LZ, a separate charcoal-based system is under development which will reduce Kr concentrations below 5~ppt for LUX, and is scalable with modest efforts. LZS requires a Kr concentration below 0.5~ppt, while LZD requires a concentration below 0.05~ppt. These levels can be achieved by scaling the LUX charcoal system.

\subsection{Neutrino Backgrounds}
\label{sub:Neutrino-Backgrounds}

LZD is expected to be the ultimate liquid xenon direct detection experiment, as the scale of the experiment expands its sensitivity to ER and NR signatures from neutrino scattering. This represents a fundamental ``noise floor'' for liquid xenon detectors, after which new technologies will be needed to further improve detection statistics. In particular, coherent neutrino scattering creates an unavoidable nuclear recoil background \cite{Strigari:2009bq}

\begin{figure}[ht!]
\centering
\includegraphics[width=80mm]{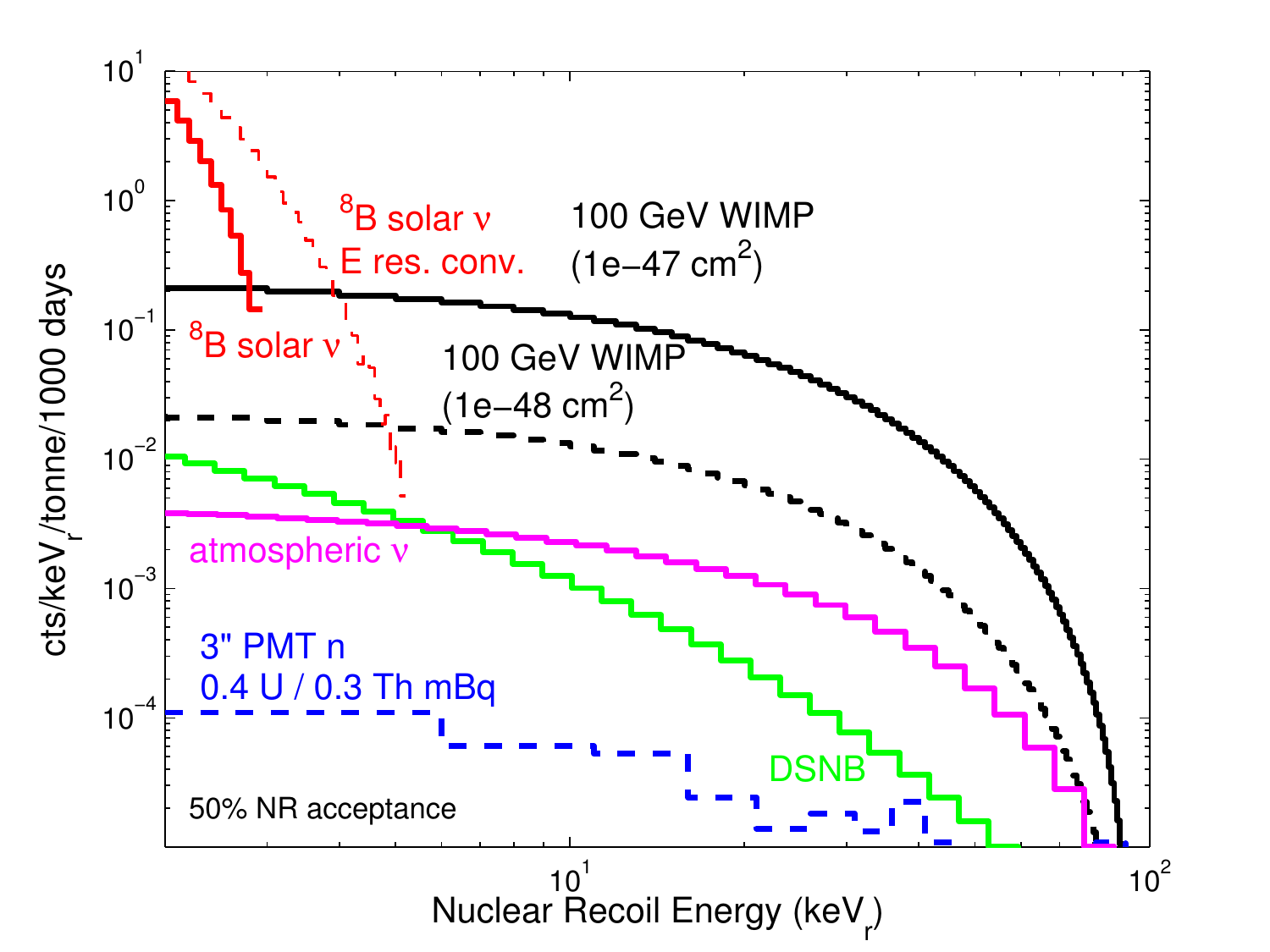}
\includegraphics[width=80mm]{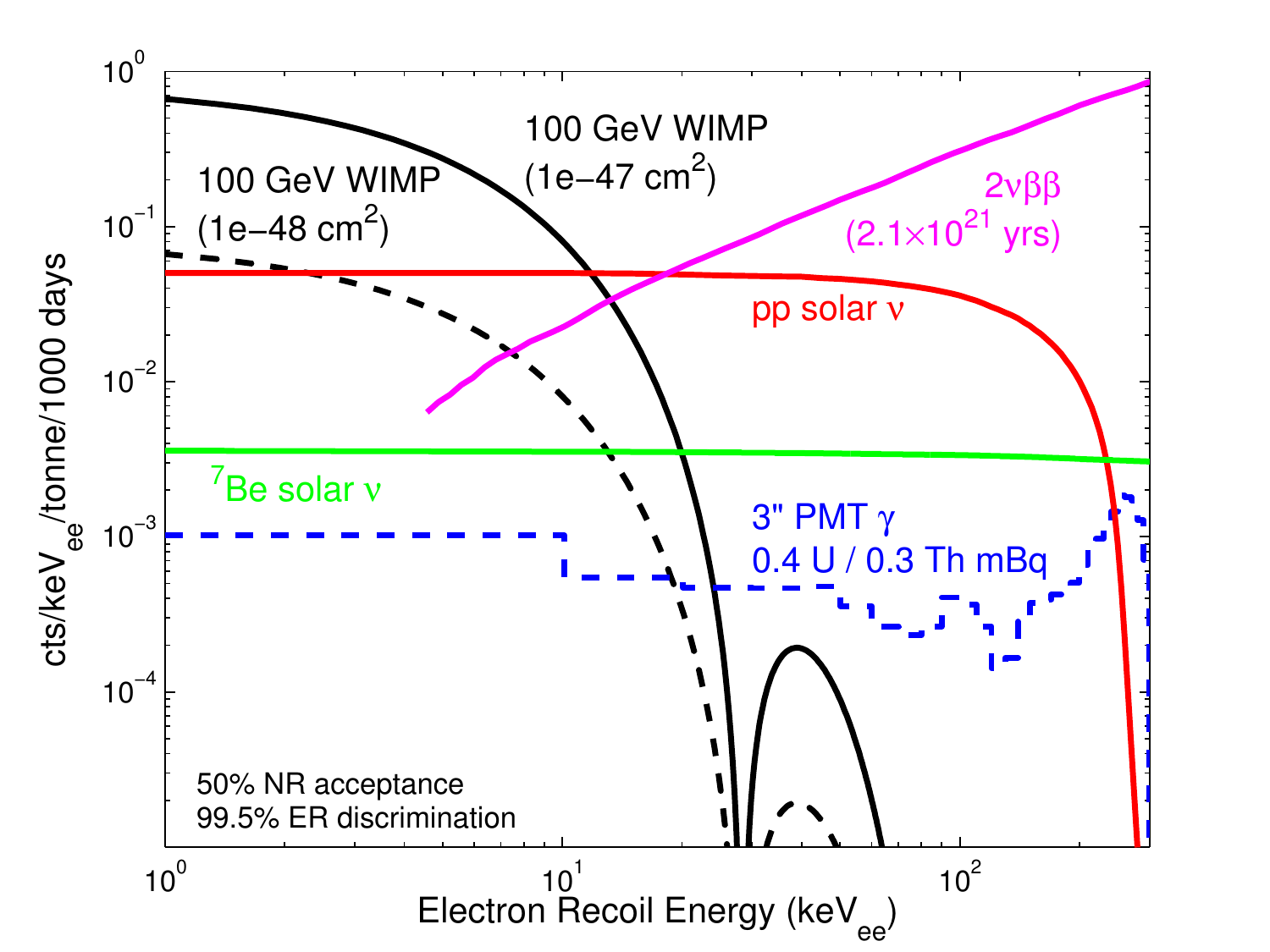}
\caption{(Left) LZD NR backgrounds from neutrino coherent scattering. The low-energy range (1-50~keV$_{\text{r}}$) where WIMP signatures are most prevalent include coherent scattering contributions from three different neutrino sources: $^8$B solar neutrinos, neutrinos from cosmic ray interactions in the atmosphere, and diffuse supernova background neutrinos. Due to the steepness of the spectrum, the $^8$B solar neutrino contribution is convolved with the expected detector energy resolution in order to estimate leakage into higher energies in the WIMP search energy range. Overlayed for comparison is the predicted neutron background spectrum for both fission and ($\alpha$,n) contributions from the PMTs, assuming the use of R11410~MOD PMTs. WIMP recoil spectra for 100~GeV particles are shown for interaction cross-sections of $10^{-47}$~cm$^2$ and $10^{-48}$~cm$^2$. A NR acceptance of 50\% is assumed. (Right) LZD ER backgrounds from p-p solar neutrinos, $^7$Be solar neutrinos, and two-neutrino double-beta decay from $^{136}$Xe, assuming a $2.1\times10^{21}$~yr lifetime as recently reported in \cite{Ackerman2011}. Overlaid are projected ER background activities from $^{238}$U and $^{232}$Th decays within the PMTs, assuming the use of R11410~MOD PMTs. A 99.5\% rejection factor is applied for neutrino and gamma spectra. Overlaid are WIMP signatures converted into the ER energy scale, assuming a conversion of keV$_{\text{r}}$/keV$_{\text{ee}}$=0.3 and a NR acceptance of 50\%.} \label{fig:LZD-NR-neutrinos}
\end{figure}

\begin{figure}[ht!]
\centering
\includegraphics[width=80mm]{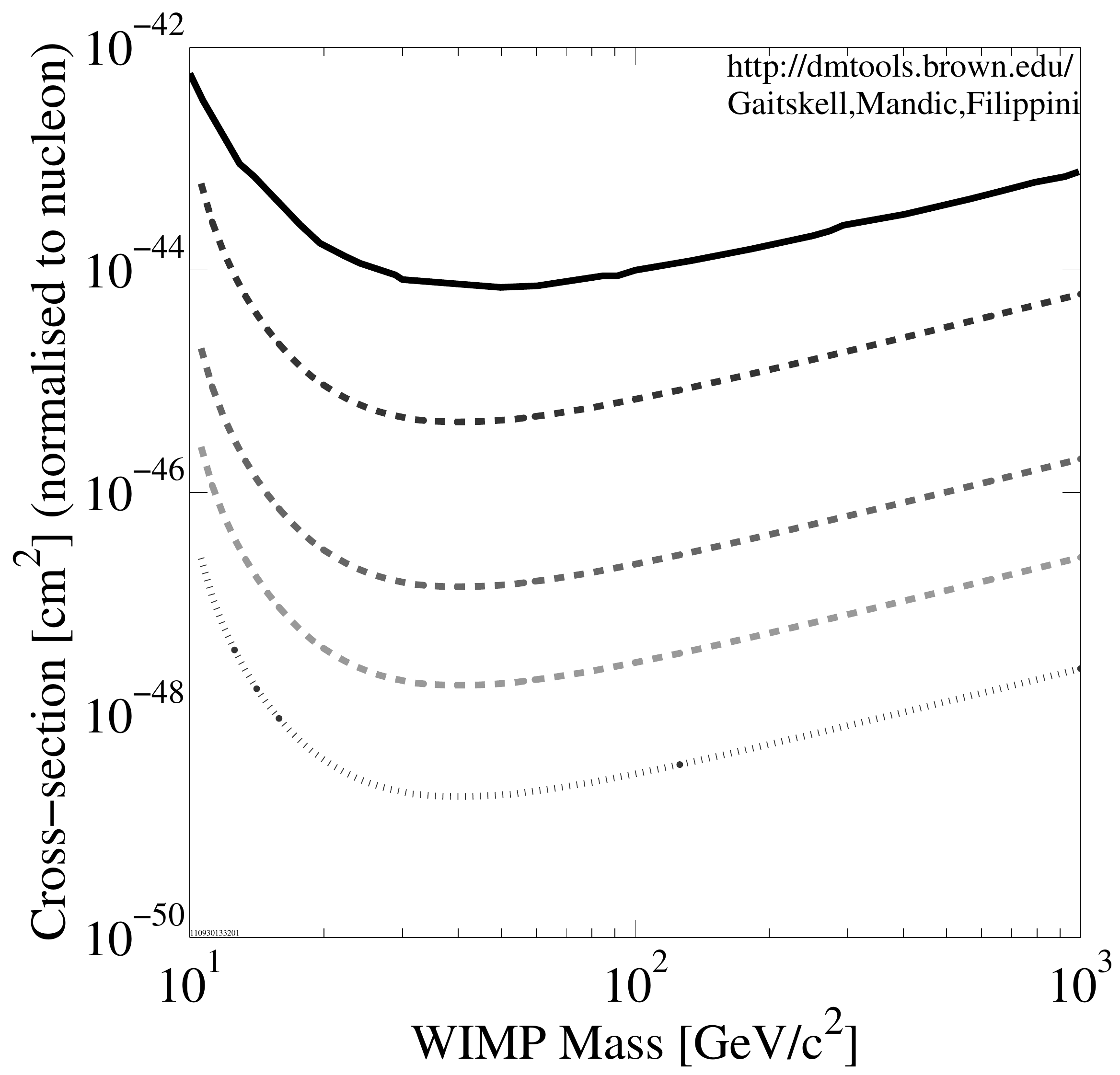}
\caption{Projected detection performance of the LUX and LZ detectors \cite{DMTools}. Dashed lines correspond in descending order to LUX (1~background event in 30,000~kg~days), LZS 1.5~tonne (1~background event in 72,000~kg~days), and LZD (10 and 1~background events for dashed and dotted, respectively, in $1.4\times10^7$~kg~days) for spin-independent interactions. Comparison is given to the lowest current limit from XENON100 (solid) \cite{Aprile2011}. } \label{fig:limit-plot}
\end{figure}

As shown in Fig. \ref{fig:LZD-NR-neutrinos} (left), the primary NR backgrounds expected for LZD come from a combination of atmospheric, solar ($^8$B), and diffuse supernova neutrino coherent scatters. Shown for reference is the activity predicted from neutrons generated from $^{238}$U and $^{232}$Th isotopes in 7.6~cm diameter PMTs, including ($\alpha$,n) backgrounds as well as $^{238}$U fission. Overlayed are predicted signatures from 100~GeV WIMPs with SI interaction cross-sections of $10^{-47}$ and 10$^{-48}$~cm$^2$; for WIMPs with lower interaction cross-sections, the WIMP signal is significantly masked by neutrino coherent scattering signatures. Electron recoils backgrounds are also limited by neutrino scattering signatures, as shown in Fig. \ref{fig:LZD-NR-neutrinos} (right). In particular, signatures from ER scatters by neutrinos from the solar p-p chain dwarf the signatures from PMT $^{238}$U and $^{232}$Th decays for energies up to 200~keV. Above 60~keV, significant contributions are present from two-neutrino double-beta decay from $^{136}$Xe, recently measured in \cite{Ackerman2011}.

\section{Projected Performance}

The projected performance for LZ WIMP detection is shown in Fig. \ref{fig:limit-plot} \cite{DMTools}. LUX projected sensitivity, assuming 1~NR background event in 30,000~kg~days, will be able to place a 90\% confidence level exclusion limit at $7\times10^{-46}$~cm$^2$ for a 100~GeV WIMP. LZS, assuming a 72,000~kg~day exposure with 1~NR background event, will improve this limit by over an order of magnitude, at $4\times10^{-47}$~cm$^2$. LZD will be able to establish a limit of $3\times10^{-48}$~cm$^2$, assuming 10~NR background events in $1.4\times10^7$~kg~days, or $5\times10^{-49}$~cm$^2$ assuming 1~NR background event for the same exposure. As mentioned in Sec. \ref{sub:Neutrino-Backgrounds}, LZD represents the limit of liquid xenon technology for WIMP detection, as signatures from neutrino coherent scattering ultimately limit these types of direct detection searches.

\begin{acknowledgments}
This work was partially supported by the U.S. Department of Energy (DOE) under award numbers DE-FG02-08ER41549, DE-FG02-91ER40688, DOE, DE-FG02-95ER40917, DE-FG02-91ER40674, DE-FG02-11ER41738, DE-FG02-11ER41751, DE-AC52-07NA27344, the U.S. National Science Foundation under award numbers PHYS-0750671, PHY-0801536, PHY-1004661, PHY-1102470, PHY-1003660, the Research Corporation grant RA0350, the Center for Ultra-low Background Experiments at DUSEL (CUBED), and the South Dakota School of Mines and Technology (SDSMT). We gratefully acknowledge the logistical and technical support and the access to laboratory infrastructure provided to us by the Sanford Underground Research Facility (SURF) and its personnel at Lead, South Dakota.
\end{acknowledgments}

\bigskip 

\end{document}